\documentclass[a4paper,fleqn]{cas-dc}

\usepackage[authoryear]{natbib}
\usepackage{amsfonts, amsmath, amssymb}
\usepackage{graphicx}	
\usepackage{amsmath}
\usepackage{listings}

\def\tsc#1{\csdef{#1}{\textsc{\lowercase{#1}}\xspace}}
\tsc{WGM}
\tsc{QE}
\tsc{EP}
\tsc{PMS}
\tsc{BEC}
\tsc{DE}

\begin{document}
\let\WriteBookmarks\relax
\def\floatpagepagefraction{1}
\def\textpagefraction{.001}

\shorttitle{ML for L\&T Brown Dwarfs Search in Sky Surveys}

\shortauthors{Aleksandra Avdeeva  et~al.}

\title [mode = title]{Machine
learning methods for the search for L\&T brown dwarfs in the data of modern sky surveys.}                      
\tnotemark[1]

\tnotetext[1]{This document is the results of the research project funded by Non-commercial Foundation for the Advancement of Science and Education INTELLECT.}

%
\author[1,2]{Aleksandra Avdeeva}[type=editor,
                        auid=000,bioid=1,
                        orcid=0000-0002-0513-4425]

\cormark[1]

\fnmark[1]

\ead{avdeeva@inasan.ru}

\credit{Conceptualization of this study, Methodology, Software, Data curation, Writing - Original draft preparation}

\affiliation[1]{organization={HSE University},
    addressline={20 Myasnitskaya St.}, 
    city={Moscow},
    postcode={101000}, 
    country={Russia}}

\affiliation[2]{organization={Institute of Astronomy RAS},
    addressline={48 Pyatnitskaya St.}, 
    city={Moscow},
    postcode={119017}, 
    country={Russia}}

\cortext[cor1]{Corresponding author}

\begin{abstract}
According to various estimates, brown dwarfs (BD) should account for up to 25 percent of all objects in the Galaxy. However, few of them are discovered and well-studied, both individually and as a population. Homogeneous and complete samples of brown dwarfs are needed for these kinds of studies. Due to their weakness, spectral studies of brown dwarfs are rather laborious. For this reason, creating a significant reliable sample of brown dwarfs, confirmed by spectroscopic observations, seems unattainable at the moment. Numerous attempts have been made to search for and create a set of brown dwarfs using their colours as a decision rule applied to a vast amount of survey data. In this work, we use machine learning methods such as Random Forest Classifier, XGBoost, SVM Classifier and TabNet on PanStarrs DR1, 2MASS and WISE data to distinguish L and T brown dwarfs from objects of other spectral and luminosity classes. The explanation of the models is discussed. We also compare our models with classical decision rules, proving their efficiency and relevance. 
\end{abstract}




\maketitle

\section{Introduction}

Brown dwarfs are substellar objects that were theoretically predicted \citep{Kumar1963,Hayashi1963} and then discovered 30 years later \citep{Rebolo1995, 1995Natur.378..463N}. Since then, the search \citep{ Luhman2013, Burningham2013, CarneroRosell2019} and systematic study of known brown dwarfs \citep{1999ApJ...519..802K, Skrzypek2016, Kirkpatrick2021} has not stopped. Their mass is insufficient to start and maintain stable hydrogen fusion, which causes them to cool over time. The peak of the radiation intensity falls into the infrared range, so objects are rather faint in the visible spectrum. In the spectral classification, brown dwarfs occupy spectral types L, T and Y. 

According to studies \citep{Muzic2017}, the number of brown dwarfs in the Galaxy ranges from 25 to 100 billion objects (with the total number of objects ranging from 100 to 500 billion). Homogeneous and complete samples of brown dwarfs are needed for various kinds of studies: kinematic studies \citep{2014MNRAS.443.2327S}, studies of binary stars with brown dwarfs \citep{2014A&A...569A.120L}, and studies of the parameters of the Galaxy. Brown dwarfs occupy the boundary between stars and planets, and studying their properties helps to refine our understanding of this boundary. Complete and uniform catalogues enable to identify and characterize brown dwarfs with greater accuracy, allowing for a better determination of the lower mass limit for stellar formation and the upper mass limit for planet formation. Moreover, brown dwarfs share similarities with giant exoplanets, making them valuable analogs for studying exoplanetary atmospheres. By studying the atmospheres of brown dwarfs, similar to exoplanets, we can gain insights into the processes and conditions that govern exoplanet atmospheres, including the presence of clouds, atmospheric composition, and thermal profiles. 

Probably the most topical issue regarding brown dwarfs is the L/T transition \citep{Artigau2009PHOTOMETRICVO, 2013A&A...555L...5G, Khandrika2013ASF}. The L/T transition in brown dwarfs is a fascinating phenomenon that is characterized by a sharp change in the near-infrared colours and brightness of brown dwarfs. It is believed to be driven by several possible mechanisms. Cloud models link the sharp transition to the sinking of dust clouds below the photosphere \citep{Marley2015, Charnay2018}. Instability in the carbon chemistry of brown dwarf atmospheres has been proposed as another mechanism contributing to the L/T transition \citep{Tremblin2019}. Adiabatic convection triggered by this instability can lead to variability across the L/T spectral sequence. Cloud dispersal has emerged as a potential mechanism for the L/T transition \citep{Tan2019}. It has been suggested that clouds with larger particle sizes dissipate more easily than clouds with smaller particle sizes. The shift from L to T spectral types may be accompanied by a change from small to large particles, leading to the fragmentation of clouds and a transition to cloud-free atmospheres \citep{Burningham2017, Saumon2008}. A detailed overview of the problem can be found in \cite{Vos2019}.

To uncover the nature of the phenomenon mentioned above, as well as to refine the statistical characteristics of brown dwarfs, complete and uniform catalogues of brown dwarfs are required. While spectroscopy is essential for confirming the nature of a brown dwarf and studying its detailed properties, conducting spectroscopic observations for a large number of objects across the entire sky is time-consuming and resource-intensive. Photometric surveys, on the other hand, can cover a much larger area of the sky and efficiently capture data on numerous celestial objects simultaneously.

By employing colour selection techniques in photometric surveys, one can identify objects that exhibit colours indicative of brown dwarf characteristics. The advantage of using photometric surveys is that they allow for systematic and wide-scale screening of potential brown dwarf candidates, helping to identify promising targets for subsequent spectroscopic observations. 

As an illustration, \cite{Skrzypek2016} accomplished this feat by effectively employing data from three surveys: SDSS, UKIDSS, and WISE. They employed a specific colour selection criterion, namely $(Y - J)_{Vega} > 0.8$ and $J < 17.5$, as a decision rule. Through this approach, they were able to discern approximately 1300 brown dwarfs within an area of 3000 square degrees, which corresponds to approximately 7.5\% of the celestial sphere. An additional noteworthy implementation in the quest for brown dwarfs is exemplified by \cite{CarneroRosell2019}. Their study also incorporated a decision rule utilizing data from the DES, VHS, and WISE surveys. The following criteria were applied: $(i - z) > 1.2$, $(z - Y) > 0.15$, $(Y_{AB}$-$J_{Vega}) > 1.6$, and $z< 22$. The imposition of a magnitude limit on the $z$ band was necessary to ensure the completeness of the dataset, thereby excluding any missing values. Within an area spanning 2400 square degrees about 5.8\% of the celestial sphere, approximately 12 thousand brown dwarfs were successfully identified through their approach.  \citet{CarneroRosell2019} also presents a comprehensive review of other colour selection works.  

When it comes to using colour selection to search for brown dwarfs, incorporating machine learning methods can provide significant advantages. Machine learning techniques can enhance the effectiveness and efficiency of the colour selection process by leveraging large datasets and complex algorithms to identify patterns and make more accurate predictions.

Machine learning methods can help uncover subtle relationships and correlations in multi-dimensional colour space, allowing for the identification of distinct colour signatures associated with brown dwarfs. This can be particularly valuable when dealing with complex and overlapping colour distributions between different objects.

ML methods are increasingly being used for classifying astronomical objects due to the vast amount of data collected in the past decades. For example, \cite{Maravelias2022} combined Support Vector Machine (SVM), Random Forest (RF) and Multilayer Perceptron to classify massive stars in nearby galaxies. The accuracy of the test dataset was 83\%. Applying it to other galaxies (not included in the dataset), namely IC 1613, WLM and Sextans A, achieved an accuracy result at the level of 70\%, which the authors attribute to different metallicity and extinction effects. The missing data were filled in with simple averages and the \textit{Iterative Imputer} method of the Scikit-learn library \citep{scikit-learn}. The Iterative Imputer method calculates the missing values based on the present values for the features in the same manner as regression models do. This method, being at the same time more robust, showed a better performance in the work.

The interpretable machine learning techniques (Localized General Matrix LVQ and RF) were used in \cite{Mohammadi2022} to detect extragalactic Ultra-compact dwarfs and Globular Clusters. Authors analysed the importance of features and compared them with features that carry physical information of the objects. 

This work aims to develop an additional tool for the search for brown dwarfs in large photometric surveys with machine learning methods. That is, based on the set of magnitudes and colours of the object, the model must determine whether the given object is a brown dwarf or not. We also compare our results with some classical decision rules: \cite{Burningham2013} and \cite{CarneroRosell2019}. The summary of these rules is shown in Tab.~\ref{tab:tab1}. The tool will be used in future work for the search for previously undiscovered brown dwarfs.

This paper is organised as follows. Sec.~\ref{sec:data} describes the dataset and preprocessing of the data, including feature engineering,  augmentation and the approach to handling the missing values. In Sec.~\ref{sec:appl} we apply the machine learning methods to the dataset and explore the importance of features. Finally, in Sec.~\ref{sec:res} we compare the performance of machine learning models and the classical decision rules and discuss the robustness of the models.

\section{Building the dataset and preprocessing}
\label{sec:data}

The dataset is based on L and T brown dwarfs from the \cite{Best2018} catalogue. The catalogue contains information on 1601 L and T type brown dwarfs and 8287 M type red dwarfs, the spectral class closest in physical characteristics to brown dwarfs. Magnitudes in 12 photometric bands and their errors are provided: $g, r, i, z, y$ of Pan-STARRS 1 \citep{Chambers2016}, $J, H, Ks$ of 2MASS space mission \citep{Cutri2003} and $W1, W2, W3, W4$ of WISE space mission \citep{Cutri2021}. 

\begin{figure*}[ht]
    \centering
    \includegraphics[width=0.9\textwidth]{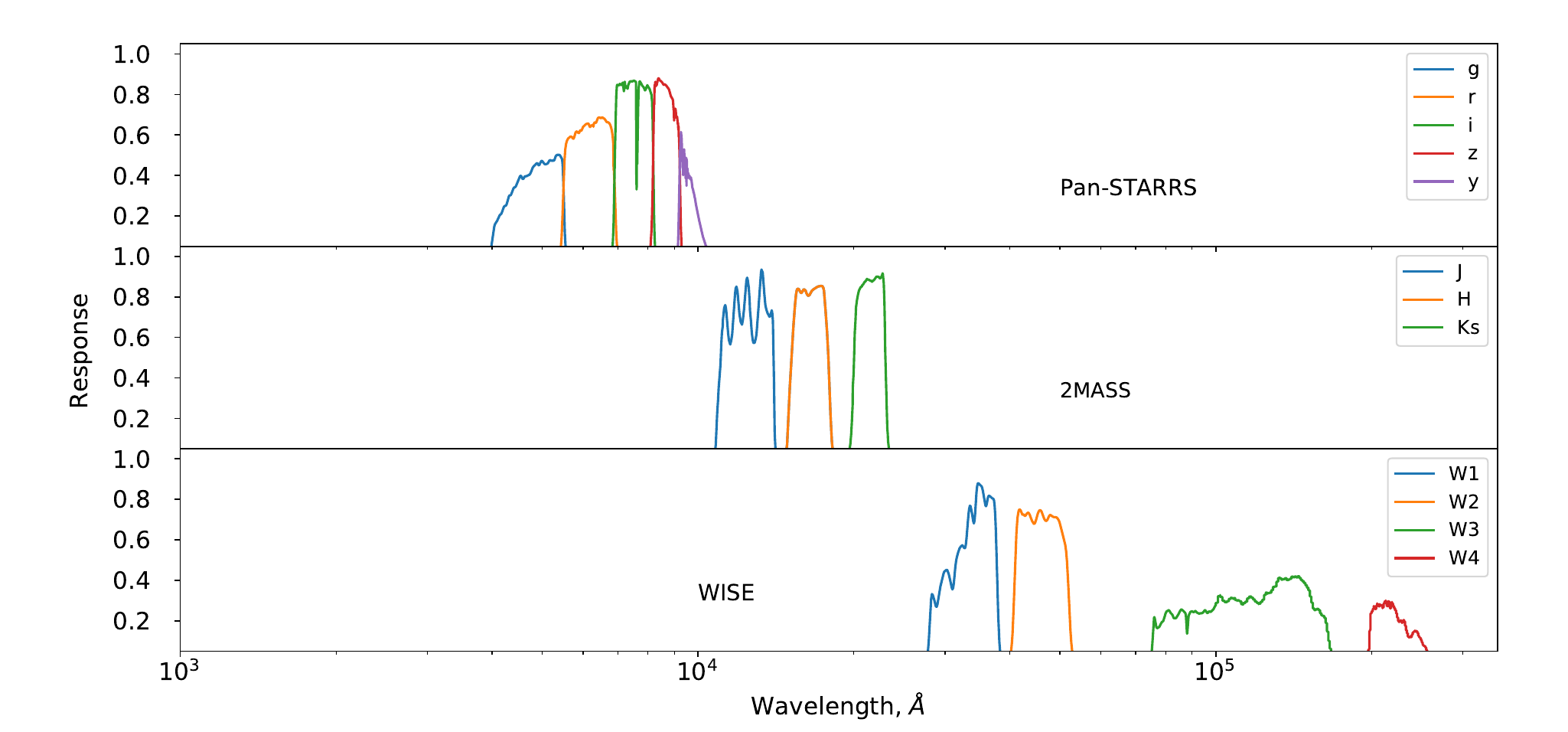}
    \caption{Response curves of photometric systems.}
    \label{fig:resp}
\end{figure*}

The catalogue also contains astrometrical information: position, parallax and proper motion. In addition, references to the literature are presented, from which data on proper motion and parallax was taken.

\begin{figure}[ht]
	\begin{minipage}[ht]{0.98\linewidth}
		\center{\includegraphics[width=\textwidth]{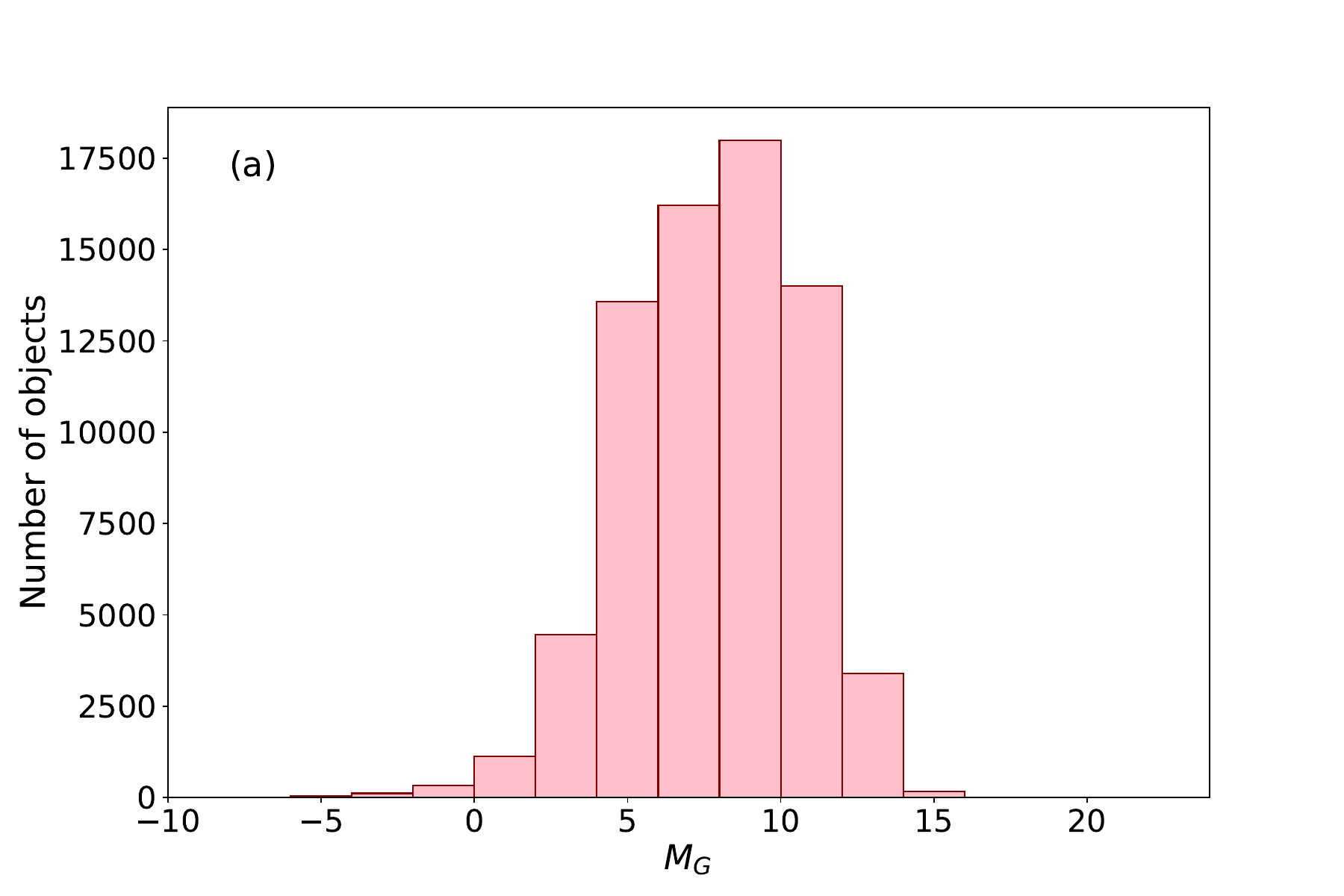}}
	\end{minipage}
	\\
	\begin{minipage}[ht]{0.98\linewidth}
		\center{\includegraphics[width=\textwidth]{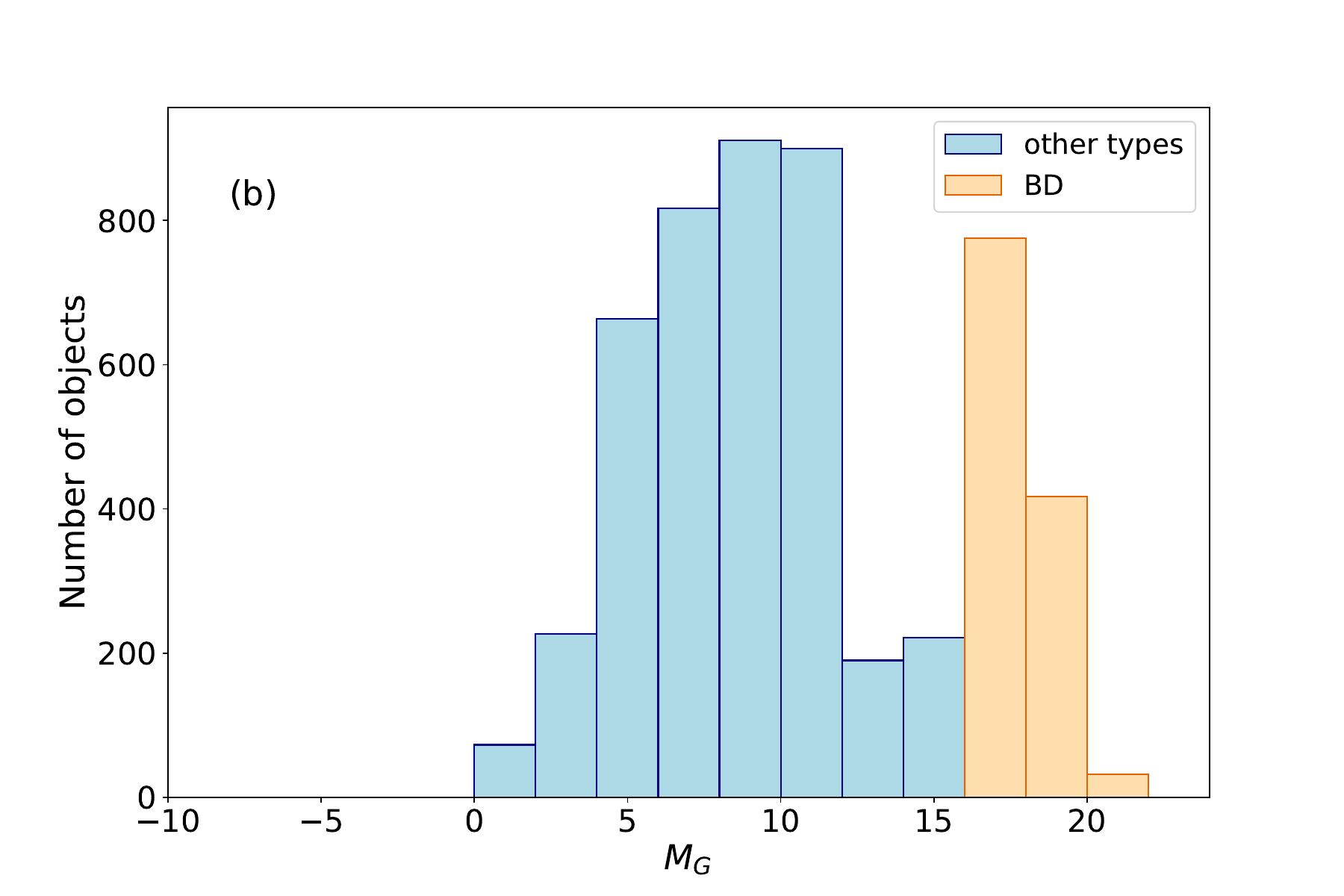}}
	\end{minipage}     \caption{Absolute colour distribution for the Gaia sample (a) and the dataset used in this work (b).} 
    \label{fig:gaia+dataset}
\end{figure}

For our Machine Learning models we consider brown dwarfs to be a positive class. To create a representative distribution of negative class objects, we examined the distribution of 100 thousand stars from Gaia DR3 \citep{Collaboration2016, Collaboration2022} by absolute magnitude $M_G$ (Fig.~\ref{fig:gaia+dataset}a). We have selected 1791 objects from A0 to K9 spectral class in the proportions observed in Fig.~\ref{fig:gaia+dataset}a from the Simbad\footnote{http://simbad.cds.unistra.fr/simbad/} database of astronomical objects. Objects that are presented in Simbad are usually well-studied and have solid spectral classifications. Gaia data seem to be short regarding M-type dwarfs, especially, after M3, so we adopt their distribution from \cite{Best2018}. The distribution of the dataset we have obtained is shown in Fig.~\ref{fig:gaia+dataset}b.

It should be noted, the obtained distribution is most likely incomplete, both in M-dwarfs and in earlier spectral classes. The deficiency in M-dwarfs could be due to \cite{Best2018} being not complete for M-type dwarfs outside of 10-pc. The objects of A0 to K9 might tend to be brighter than the actual population and in fact have an underdensity in the range between $12^m$ and $15^m$ in $i_{PS1}$ (Fig~\ref{fig:data}a), which we, however, assume to be not relevant in case of using only colour indices as features.

The objects selected from Simbad were cross-matched with data from the Pan-STARRS DR1, 2MASS, and ALLWISE catalogues. We chose the matching radius to be $1''$, which is a reasonable value for most surveys for objects with low proper motions, including those used in this paper. 

The resulting dataset contains 5669 objects, 1601 of which are of the positive target class. The dataset is available online\footnote{https://github.com/iamaleksandra/ML-Brown-Dwarfs/}. As the peak intensity of brown dwarfs falls in the infrared part of the spectrum, their magnitudes in the optical photometric bands (bands $g$,$r$,$i$) are most likely almost beyond the sensitivity limit of the telescope and therefore are missing from the data. The $g$ and $r$ bands of the Pan-STARRS data are missing for almost all objects, so we don't use these magnitudes as features. The $i$ band values are missing for about one-third of the objects of the positive class and a small number of objects of the negative class. Magnitude values in this band are important to us, also for comparison with the classical rules, so we keep them. We also remove $W3$ and $W4$ as they are of poor quality, 90th percentile of the error of magnitude measurement for both magnitudes is about half of a magnitude, which is very noisy for the classification problem we want to solve. As a result, we have 7 magnitudes for each object: $i_{PS1}$, $z_{PS1}$, $y_{PS1}$, $J$, $H$, $Ks$, $W1$, $W2$.

\subsection{Train, validation and test}
\label{subsec:tvt}
The data is divided into training (60\%), validation (20\%) and test (20\%) sets in a stratified fashion using using sci-kit learn \textit{train\_test\_split} method. The features are scaled using \textit{StandardScaler} and model hyperparameters are selected on the validation set using \textit{optuna} \citep{Akiba2019}. The final performance of the model is calculated on the test set.

\subsection{Augmentation}
\label{sec:aug}

As the number of objects of negative class is almost 2.5 times higher than the number of positive class objects, we use augmentation with Gaussian-distributed noise as an oversampling method to make the dataset more balanced. The data was divided into training, validation and test subsamples prior to augmentation in order to not let models see the augmented data in the test sample, while being trained on the original prototype of these augmented objects. 

We augment the data of positive and negative classes separately, using all objects of positive class and only objects in the range between $12^m$ and $15^m$ in $i_{PS1}$ in negative class objects. For each feature error, we calculate the mean and the standard deviation. We then generate normally distributed noise values with the same parameters of the distribution. The noise is added to all values of the corresponding features, which do not have any missed values. Thus we have 8364 objects of which 4155 are positive and 4209 are negative.   

\subsection{Feature engineering}
\label{sec:genp}

In a classification of astronomical objects, as in various types of astronomical problems, the colours of objects are even more important than the magnitudes. Colours are characteristic of the energy distribution in the spectrum and are almost independent of distance. To take this into account when classifying, we have added several features - colour indices: $(i-z)_{PS1}$, $(i-y)_{PS1}$, $(z-y)_{PS1}$, $z_{PS1}-J$, $y_{PS1}-J$, $J-H$, $H-Ks$, $Ks-W1$, $W1-W2$. They are also often used as the colour selection to distinguish brown dwarfs from other objects. We only use colours of the most spectroscopically adjacent magnitudes (see Fig.~\ref{fig:resp}). The two exceptions are $z_{PS1}-J$ since it is commonly used as a colour selection and proved to be useful, and $(i-y)_{PS1}$ since it turned out to be extremely useful for classification (see Sec.~\ref{sec:res} for details). 

\begin{figure}[ht]
	\begin{minipage}[ht]{0.98\linewidth}
		\center{\includegraphics[width=\textwidth]{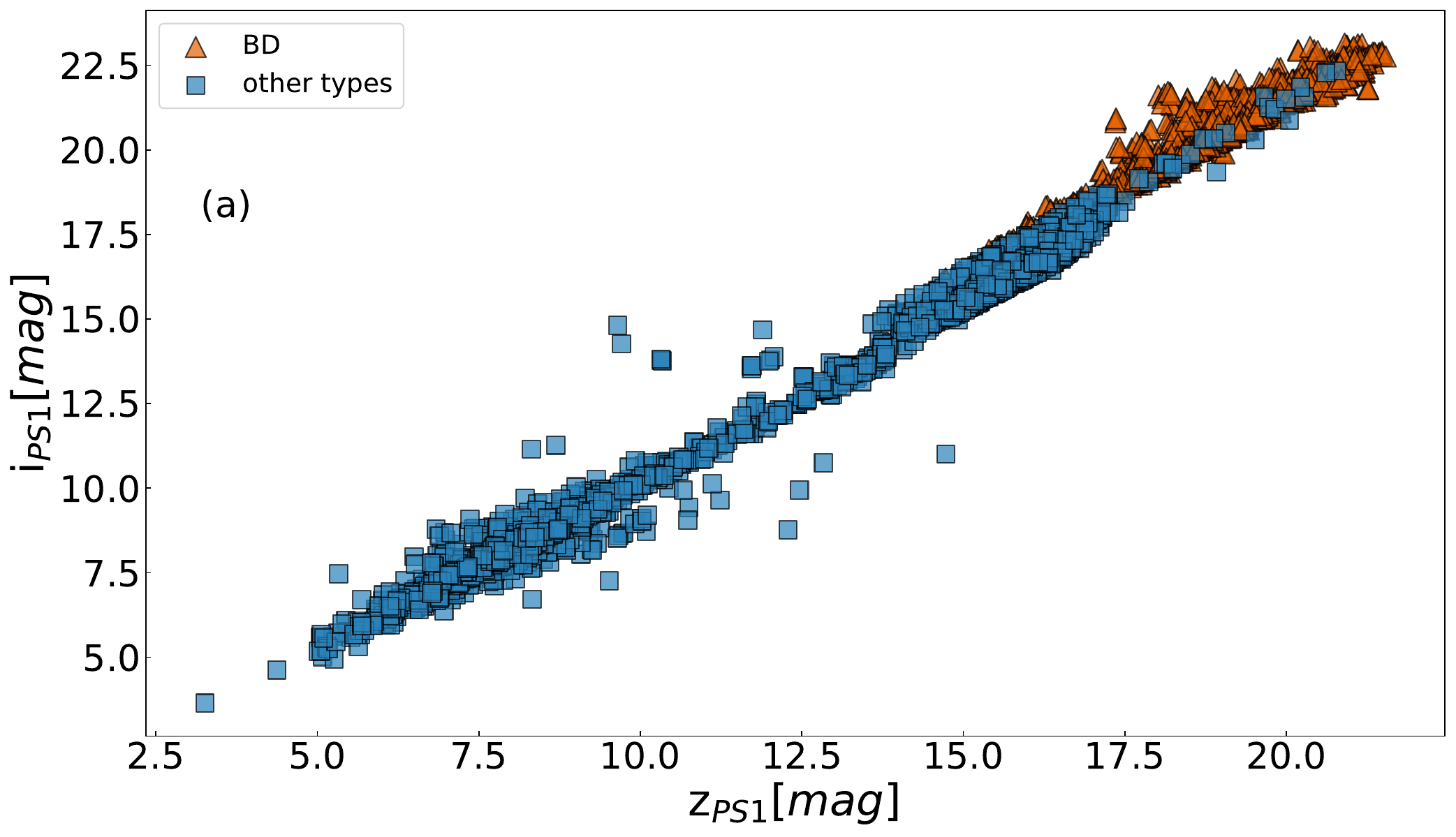}}
	\end{minipage}
	\\
	\begin{minipage}[ht]{0.98\linewidth}
		\center{\includegraphics[width=\textwidth]{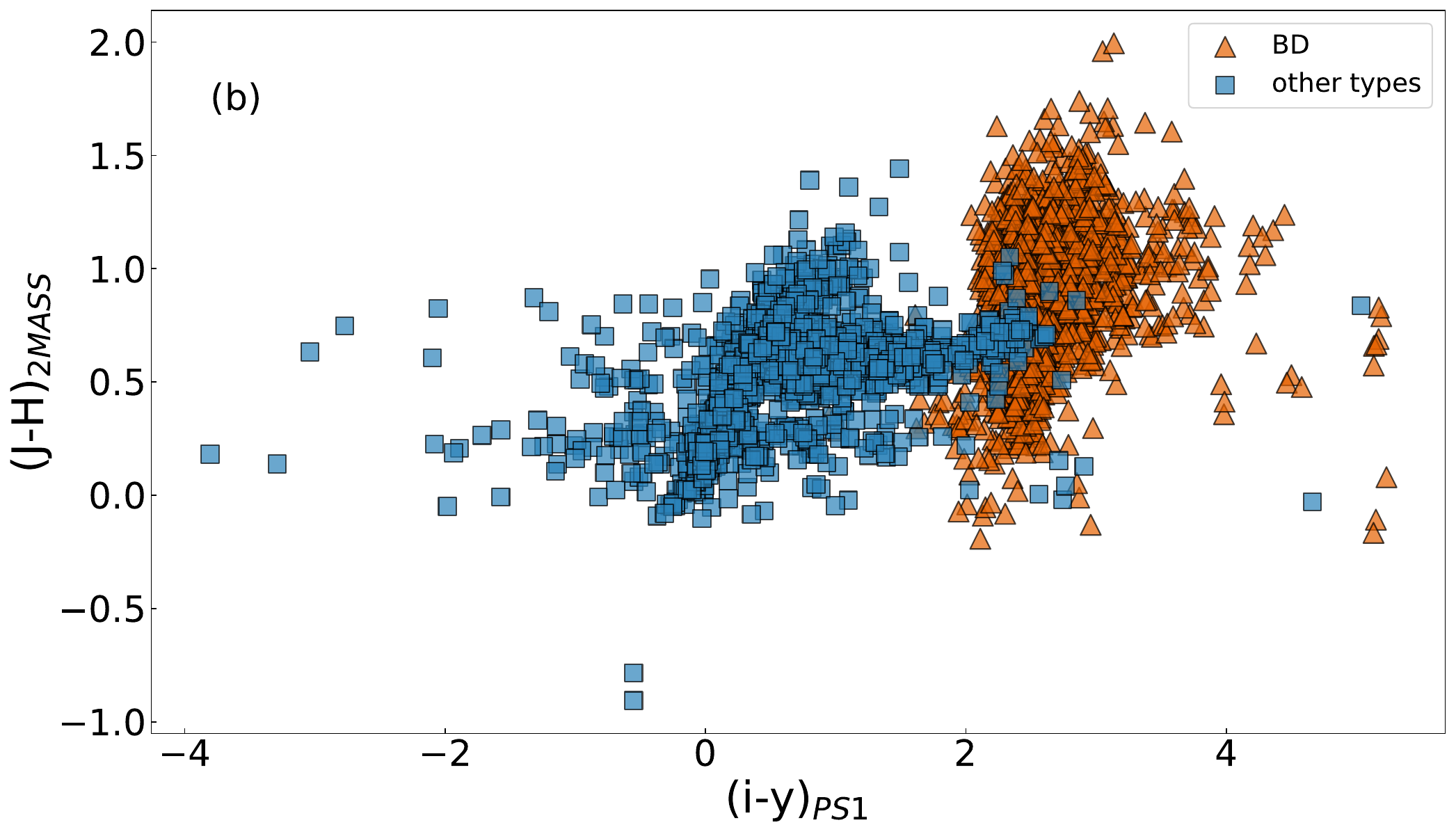}}
	\end{minipage}     \caption{Augmented data: objects of different classes on the magnitude-magnitude (a) and colour-magnitude (b) diagrams. Brown triangles are the objects of positive class (L\&T brown dwarfs), blue squares are the objects of negative class. Here blue points are plotted on top of the brown triangles.} 
    \label{fig:data}
\end{figure}

After this procedure, the table contains 17 features for each of the 8364 objects, listed in Tab.~\ref{tab:tab0}. Fig.~\ref {fig:data} shows how objects of the target class look compared to objects of all other classes in a two-dimensional slice of feature space. We use all of the magnitudes and colours simultaneously (even though the latter ones are the linear combination of the first ones by design) since we process the missing values independently, which sometimes violates these relations (for the details see Sec.~\ref{subsec:miss}).   

As one can see (Fig.~\ref{fig:data}a), the upper limit of magnitude $i_{PS1}$ differs a lot for our positive and negative objects. This is due to the procedure of building the dataset, i.e. Simbad has four times more objects with $i_{SDSS}>20$ and a spectral type $\geq L0$ than objects with a spectral type $<L0$. Catalogues, however, contain large amount of faint objects, the majority of which are not brown dwarfs. We therefore should avoid models based primarily on PS1 magnitudes. Although our dataset can be called balanced regarding other magnitudes (from 2MASS and WISE), the magnitude is not only a function of the luminosity of an object, but also a function of the distance to it, so it is not advisable to rely on these magnitudes as well. 

Thus, three cases for each approach are investigated: all magnitudes and colours are used as features (we call this case "all features"), no Pan-STARRS magnitudes are used ("w/o PS magnitudes") and no magnitudes used at all ("only colours").

\subsection{Processing of missing values}
\label{subsec:miss}

As it was mentioned earlier, the dataset has a significant number of missing values. At the part of the spectrum with longer wavelength, this is most likely due to the sensitivity limit of the telescope: brown dwarfs are rather faint objects, and their emission maximum falls into the infrared part of the spectrum. Missing values at the shorter wavelength part of the spectrum seem to occur due to poor-quality measurements or artefacts.

To deal with the missing values, \cite{Maravelias2022} used imputing by means and the Iterative Imputer of the Sckit-learn library, showing that Iterative Imputer provides results more robust and effective in terms of classification.

We test the method by additionally throwing out the magnitude values for 5 percent of objects, which do not have missing values of the particular feature. Then we impute these values using the Iterative Imputer and compare the results to the original feature values for the object. 

Tab.\ref{tab:tab0} shows the results of the imputation with the following parameters of the Iterative Imputer:

\begin{lstlisting}[language=Python]
estimator=ExtraTreesRegressor
(n_estimators=150
max_features=14
max_depth=15
min_samples_split=12
initial_strategy='median'
max_iter=20)
\end{lstlisting}

Tab.~\ref{tab:tab0} contains the information about the fraction of missing values of a particular feature in the dataset and the number of objects that were withheld for the testing. We also compare the 90th percentile error of the feature measurement (the error of magnitude is usually given in the catalogue and the error of the colour index is calculated as the square root of the sum of the squared errors of the magnitudes involved) with the 90th percentile discrepancy in the actual value of the feature and the value predicted by the imputer.  

In most cases, the 90th percentile for the discrepancy between the imputed values and the original ones is compatible with the 90th percentile for the error of the feature. Even though the colour values calculated in the Sec.~\ref{sec:genp}  are directly related to the magnitude values, it was decided to apply Iterative Imputer to them independently. This allows one to achieve better results and avoid large errors in the calculation of colour values, as can be seen from Tab.~\ref{tab:tab0}. 

\begin{figure}[ht]
	\begin{minipage}[ht]{\linewidth}		\center{\includegraphics[width=0.95\linewidth]{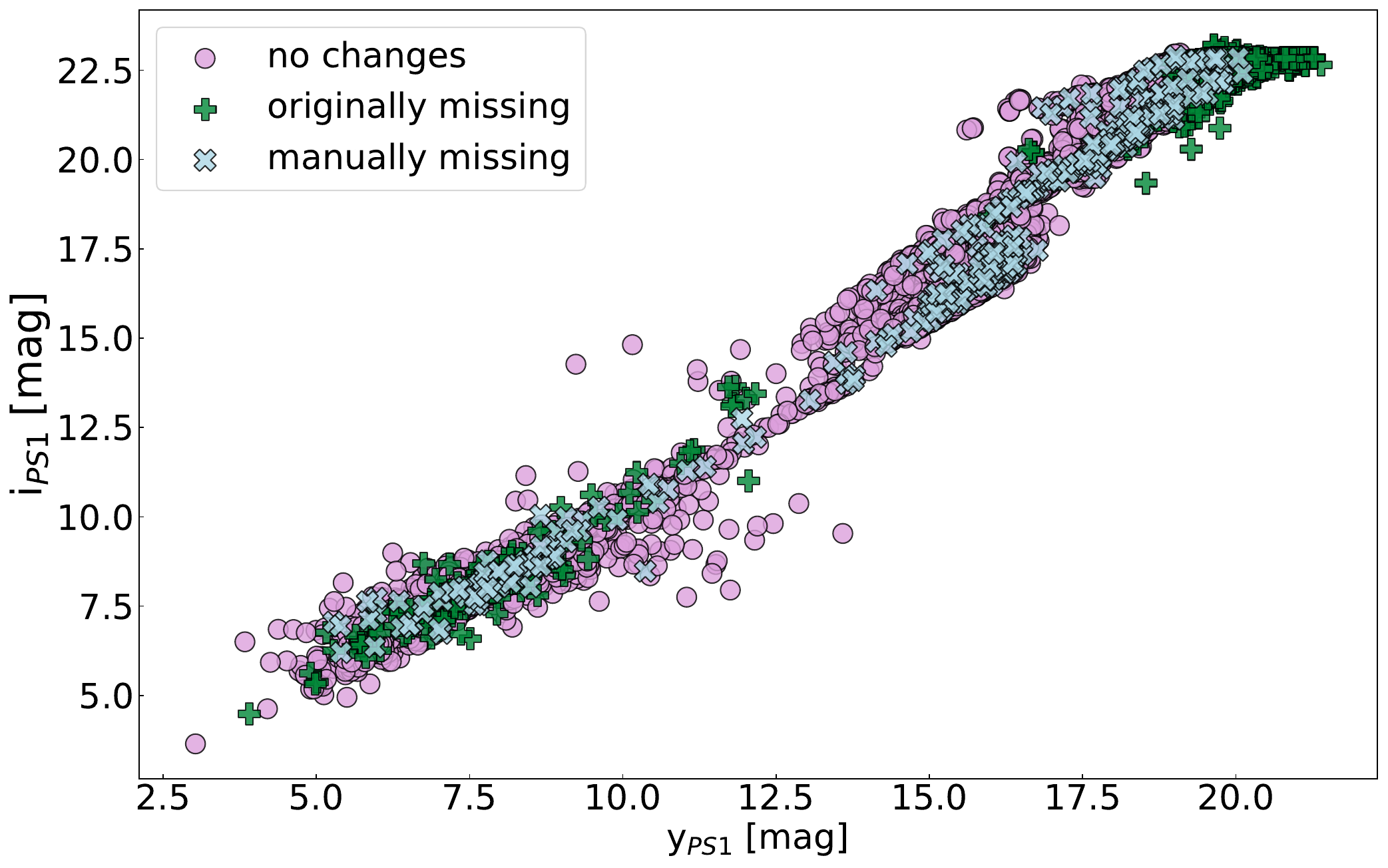}}
	\end{minipage} \\
	\vfill
	\begin{minipage}[ht]{\linewidth}		\center{\includegraphics[width=0.95\linewidth]{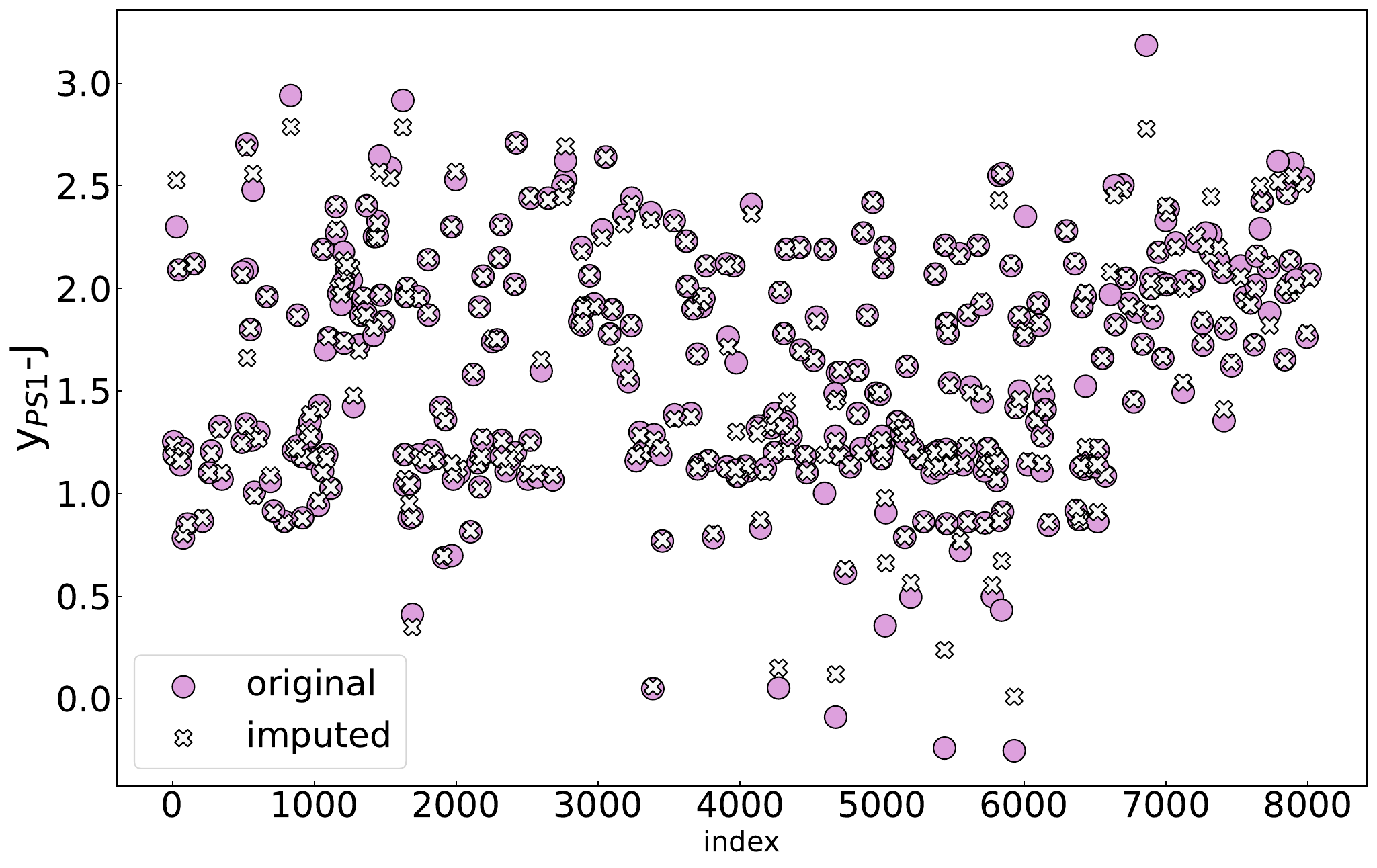}}
	\end{minipage} 
    \caption{An example of imputed values for y$_{PS1}-$J and y$_{PS1}$ features. (Top) Original data (pink circles), initially missing data (green plus) and manually missing values (blue crosses) on the colour-magnitude diagram.   (Bottom) Comparison of the original data (pink circles) and the values, imputed via the Iterative Imputer method (white crosses). An index is used to avoid contamination from the imputed values of other features.} 
    \label{fig:eximput}
\end{figure}

In Figure~\ref{fig:eximput}, we can see an example of imputing. The top panel displays a magnitude-magnitude diagram containing both the original and imputed data. It's worth noting that a significant portion of the missing data points are located in the fainter range of the $i$ magnitude. The bottom panel compares the original data to the imputed values for the same objects. Although there are some discrepancies up to $0.5^\textrm{m}$ in the $y_{PS1}-J$ color, most of the values are predicted accurately. Specifically, for 90\% of the stars, $y_{PS1}-J$ value have an error less than $0.063^\textrm{m}$.

\begin{table*}[ht]
\begin{center}
\caption{Properties of the dataset and the results of the imputing test for every feature for the train part of the dataset. The table includes the fraction of missing values in the dataset (after augmentation) and the number of objects which were withheld for the test (5\% of the objects for which the value of the feature was presented). We compare the 90th percentile for the discrepancy between the imputed values and the original ones with the 90th percentile for the error in the measurement of the feature value. The error of the colour index is calculated as the square root of the sum of the squared errors of the magnitudes involved.}
\begin{tabular}{ccccc}
Feature     & Fraction of  & Number of  & 90th percentile  & 90th percentile \\
& missing values & objects withheld & of the error & of discrepancy \\
    \hline
    \hline
     i$_{PS1}$ & 17\% & 208 & 0.050 & 0.070 \\ 
     \noalign{\smallskip}\hline\noalign{\smallskip}
     
     z$_{PS1}$ & 5.5\% & 237 & 0.050 &  0.091 \\ 
     \noalign{\smallskip}\hline\noalign{\smallskip}
     
     y$_{PS1}$ & 2.2\% & 245 & 0.060 & 0.109  \\
     \noalign{\smallskip}\hline\noalign{\smallskip}
     
     J & 8.8\% & 228 & 0.100 & 0.101  \\
     \noalign{\smallskip}\hline\noalign{\smallskip}     
     
     H & 8.8\% & 228 & 0.120 & 0.107  \\
     \noalign{\smallskip}\hline\noalign{\smallskip}     
     
     Ks & 8.9\% & 228 & 0.110 & 0.075  \\
     \noalign{\smallskip}\hline\noalign{\smallskip} 
     
     W1 & 2.1\% & 245 & 0.141 &  0.085 \\
     \noalign{\smallskip}\hline\noalign{\smallskip}  
     
     W2 & 2.0\% & 242 & 0.080 &  0.096 \\
     \noalign{\smallskip}\hline\noalign{\smallskip}  
     
     (i-z)$_{PS1}$ & 18.4\% & 204 & 0.067 &  0.050 \\
     \noalign{\smallskip}\hline\noalign{\smallskip} 
     
     (i-y)$_{PS1}$ & 18.0\% & 205  & 0.072 &  0.038 \\ 
     \noalign{\smallskip}\hline\noalign{\smallskip}  
     
     (z-y)$_{PS1}$ & 6.7\% & 234  & 0.073 &  0.077  \\ 
     \noalign{\smallskip}\hline\noalign{\smallskip}  
     
     z$_{PS1}$-J & 12.3\% & 219  & 0.122 & 0.048  \\ 
     \noalign{\smallskip}\hline\noalign{\smallskip}   
     
     y$_{PS1}$-J & 10.9\% & 223  & 0.130 & 0.063 \\ 
     \noalign{\smallskip}\hline\noalign{\smallskip}   
     
     J-H & 8.8\% & 228 & 0.164 &  0.154 \\ 
     \noalign{\smallskip}\hline\noalign{\smallskip} 
     
     H-Ks & 9.1\% & 225 & 0.170 &  0.145 \\ 
     \noalign{\smallskip}\hline\noalign{\smallskip}  

     Ks-W1 & 9.9\% & 226 & 0.183 &  0.143 \\ 
     \noalign{\smallskip}\hline\noalign{\smallskip}  
     
     W1-W2 & 2.2\% & 245 & 0.168 &  0.176 \\ 
     \hline
    \hline

\end{tabular}
\label{tab:tab0}
\end{center}
\end{table*}

\section{Application of machine learning}
\label{sec:appl}

Four approaches are tested during the work: Random Forest (RF), Support Vector Machines (SVM), XGBoost and TabNet. As it was said in Sec.~\ref{sec:data}, we investigate three cases for each approach: "all features", "w/o PS magnitudes" and "only colours". We calculate the score and feature importance for each model, using \textit{SHAP} \citep{NIPS2017_7062} for RF, XGBoost, SVM and TabNet. Although TabNet has built-in capabilities of calculating importance of features,based on the attention mechanism's dynamic selection of input features, we use \textit{SHAP} as well so as we could compare the results properly. 

The \textit{SHAP} method works by evaluating the model's prediction for each instance while permuting the values of a specific feature. This involves shuffling the values of the chosen feature while keeping the rest of the features unchanged. The difference between the model's prediction with the original feature values and the prediction with the permuted feature values is used to calculate the Shapley value. 

We compare all models to the classical decision rules \cite{CarneroRosell2019} and \cite{Burningham2013}. The Matthews correlation coefficient (MCC) is chosen as the primary metric since it takes into account both False Positive and False Negative predictions. It can be calculated as follows:

\small
\begin{equation*}
    \textsc{MCC} = \frac{\textsc{TP} \times \textsc{TN}-\textsc{FP} \times \textsc{FN}}{\sqrt{(\textsc{TN}+\textsc{FN})(\textsc{TN}+\textsc{FP})(\textsc{TP}+\textsc{FP})(\textsc{TP}+\textsc{FN})}}
\end{equation*}
\normalsize
where TP is the number of true positives, TN the number of true negatives, FP the number of false positives and FN the number of false negatives. 
We also provide the precision and recall scores for they are more intuitive. Precision is a proportion of relevant instances among retrieved instances, while recall is the proportion of relevant instances that have been retrieved. They are defined as:

\small
\begin{equation*}
    \textsc{Precision} = \frac{\textsc{TP}}{\textsc{TP}+\textsc{FP}} \\
    \textsc{Recall} = \frac{\textsc{TP}}{\textsc{TP}+\textsc{FN}}    
\end{equation*}

\normalsize

The classical decision rules and the result of their application to the test dataset after augmentation and imputation are summarized in Tab.~\ref{tab:tab1}. The MCC score was calculated on the test part of the dataset. It should be mentioned, that even though the filters in different surveys have similar names (i.e. $y_{PS1}$ and $Y_{DES}$), they are not identical to each other. Therefore, it is not entirely correct to apply the rules made for one survey to the magnitudes of the other survey. However, we have estimated that for our dataset the magnitudes of the same name differ within $0.2$ mag which does not increase the score in any case. Also, it is worth mentioning that \cite{Burningham2013} was originally devoted to T-type dwarfs exclusively, yet it shows great performance on L dwarfs, so we use it as a decision rule for both L and T dwarfs. 

Although the performance of the decision rules is reasonably high, the actual number of false positive and false negative classifications grows with the number of objects, and this becomes important when we have millions of objects, as in most modern sky surveys (PanSTARRS - 1.9 billion objects, 2MASS - 470 million objects, ALLWISE - 560 million objects), so it makes sense to make an effort to increase the performance and these values (Tab.~\ref{tab:tab1}) are the baselines we want to outperform.

\begin{table*}[ht]
\begin{center}
\caption{Decision rules from the literature.}
\begin{tabular}{c|c|c}
Author     & Rule      & MCC        \\
\noalign{\smallskip}\hline\noalign{\smallskip}
     Carnero Rosell et al. (2019) & $(i - z) > 1.2$, $(z - Y) > 0.15$, &
     0.935\\
     & ($Y_{AB}-J_{Vega}) > 1.6$, z< 22 & \\
     \noalign{\smallskip}\hline\noalign{\smallskip}
     Burningham et al. (2013) & $(z - J)_{Vega} > 2.5$, $J < 17.5$ & 0.921
\end{tabular}
\label{tab:tab1}
\end{center}
\end{table*}

\begin{figure}[ht]
	\begin{minipage}[ht]{\linewidth}
		\center{\includegraphics[width=0.7\textwidth]{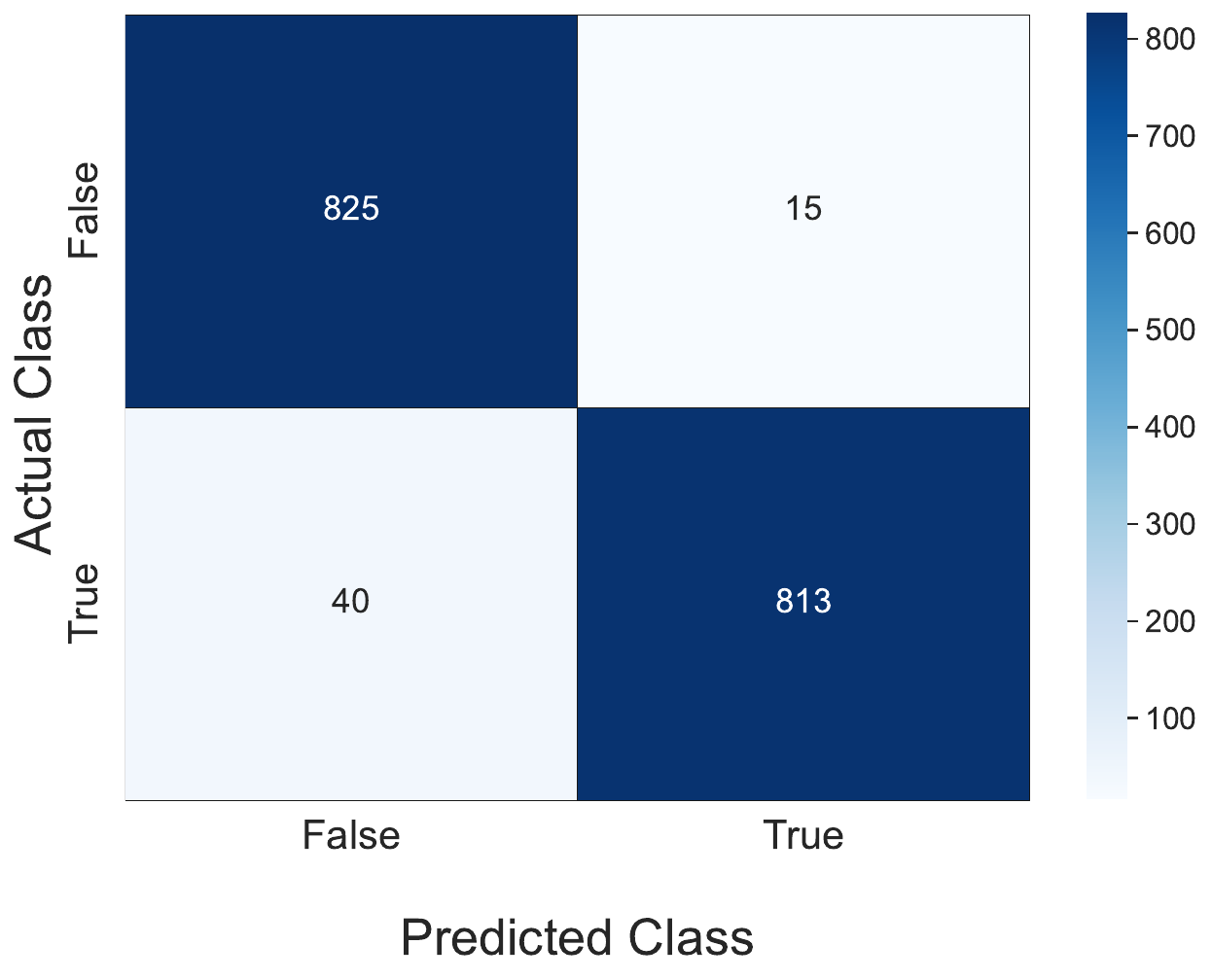}}
	\end{minipage} 
        \\
	\begin{minipage}[ht]{\linewidth}	\center{\includegraphics[width=0.7\textwidth]{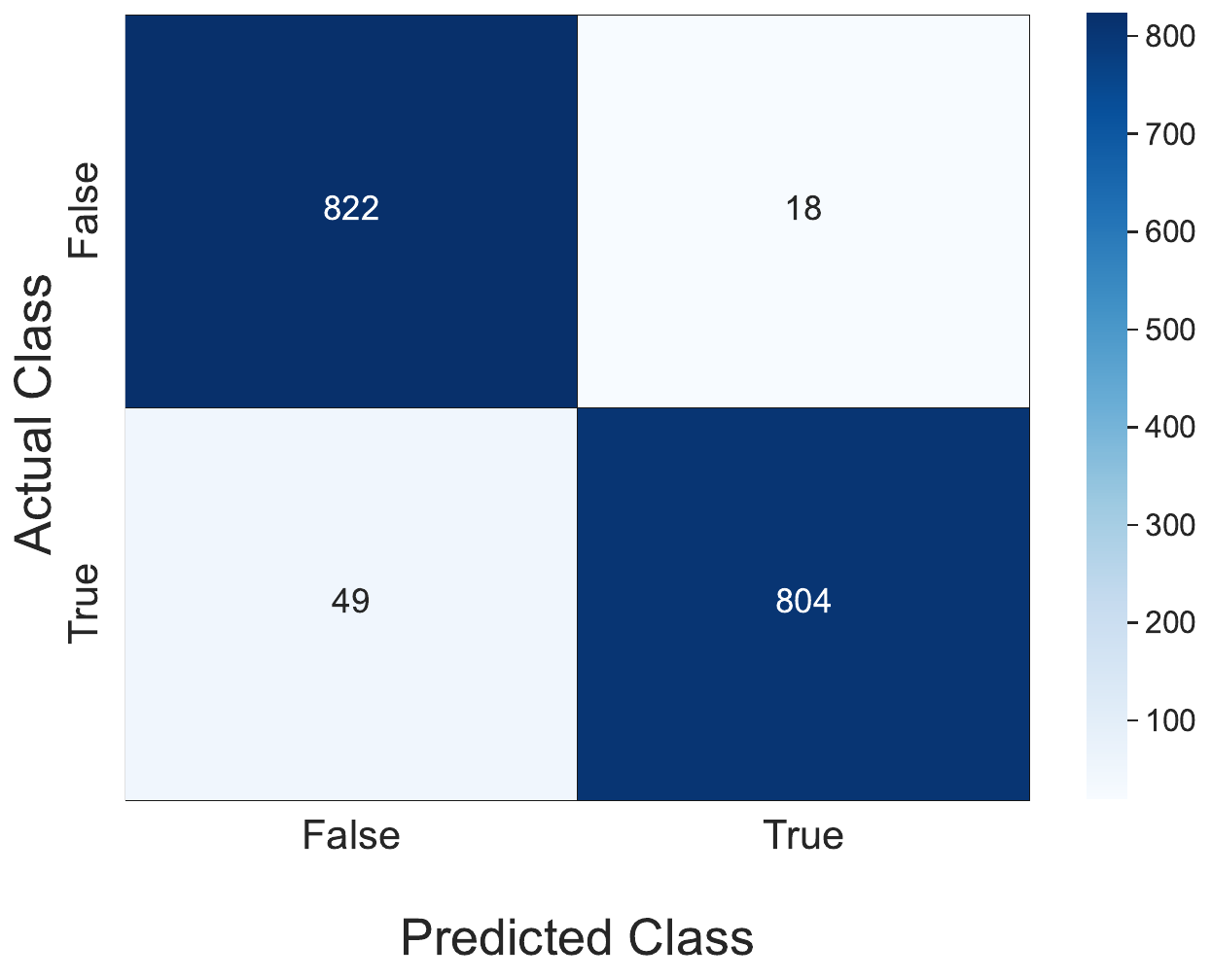}}
	\end{minipage} 
    \caption{The confusion matrices for the decision rules \cite{CarneroRosell2019} on the top and \cite{Burningham2013} on the bottom applied to the test part of the dataset.} 
    \label{fig:cm_dr}
\end{figure} 

The work made use of the following Python software packages: Sckit-learn 1.0.2 version, optuna 2.10.1 version, Tabnet from Pytorch 4.0 version.

\subsection{Random Forest}

The decision tree concept is quite similar to the classical colour selections that are traditionally used in astronomy to classify objects. While automated decision trees can be much more efficient than classic decision rules, they tend to overfit, i.e. they learn too much about the data they are trained on and can fail when applied to data they have not seen before. The solution to this problem can be a random forest approach (RF) - an ensemble of decision trees. In this case, the decision about the class an object belongs to, is made based on which class the greater number of trees voted for.

Using \textit{optuna}, we selected the maximum tree depth, the minimum number of samples required to split an internal node, the criterion and the maximum number of features in a node. Tuned parameters and corresponding scores are presented in Tab.~\ref{tab:tab_rf}. Note that the scores are not guaranteed and depend on the dataset the model is tested on, e.g. the imputing of missing values. For a more objective evaluation of the model we use bootstrap technique (see confidence intervals, Sec.~\ref{sec:res} and Fig.~\ref{fig:trust}).

\begin{figure}[ht]
	\begin{minipage}[ht]{\linewidth}	\center{\includegraphics[width=0.8\textwidth]{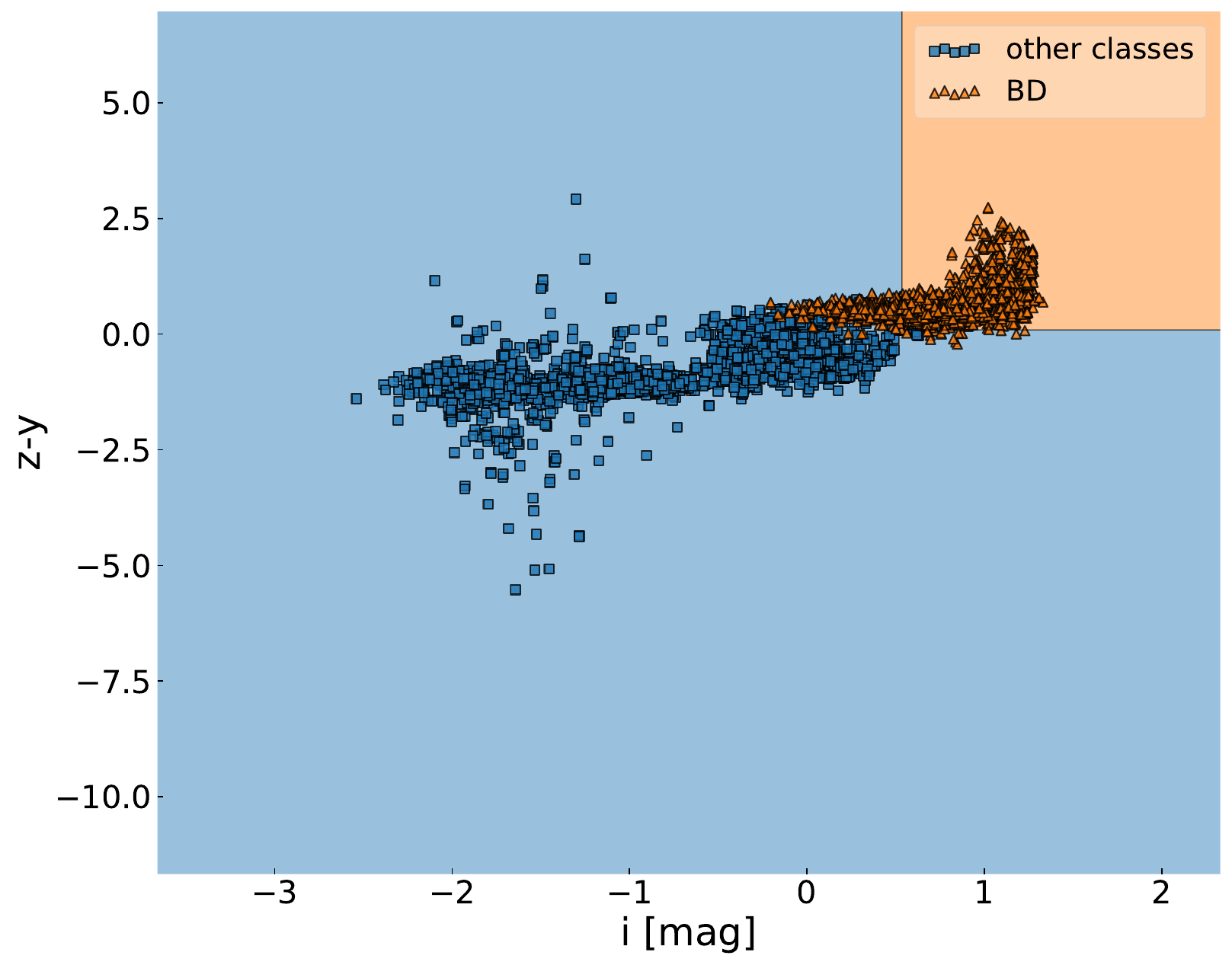}}
	\end{minipage}
    \caption{Slice of the separating boundary in the feature space according to the RF model.}
    \label{fig:cut_rf}
\end{figure}

In Fig.~\ref{fig:cut_rf}, an example of a two-dimensional cut shows the separating boundary between classes defined by the RF model. It should be noted that this is a slice in multidimensional parameter space, so it does not represent the number of mismatches. In fact, the performance is very good, meaning that the decision rules underlying that model might be quite complex and cannot be represented on a 2D diagram.

Fig.~\ref{fig:importance} shows the importance of features for the RF model in all three cases. Seeing Pan-STARRS magnitudes, the model primarily relies on $i$ magnitudes. Without optical magnitudes, $i-y$ becomes the most important feature. This becomes even more pronounced for the RF model that relies on colours only. 

\begin{table}[ht]
\begin{center}
\caption{Random forest hyperparameters for three sets of features. Number of trees is 500 for all of the models. Number in max features is the fraction of all available features.}

\scriptsize
\begin{tabular}{lccc}
Hyperparameter     & All features      & No PS magnitudes & Only colours  \\
\hline
\hline
Max depth & 11 & 13 & 12 \\
\noalign{\smallskip}\hline\noalign{\smallskip}
Min samples split & 20 & 9 & 8 \\
\noalign{\smallskip}\hline\noalign{\smallskip}
Max features & sqrt & sqrt & 0.7 \\
\noalign{\smallskip}\hline\noalign{\smallskip}
Criterion & entropy & entropy & gini \\
\hline
\hline
\noalign{\smallskip}\noalign{\smallskip}
MCC test score & 0.983 & 0.986 & 0.975 \\
\noalign{\smallskip}\hline\noalign{\smallskip}
MCC train score & 0.987 & 0.990 & 0.990 \\
\noalign{\smallskip}\hline\noalign{\smallskip}
Precision & 0.992 & 0.992 & 0.988 \\
\noalign{\smallskip}\hline\noalign{\smallskip}
Recall & 0.992 & 0.994 & 0.987 \\
\hline
\hline
\end{tabular}
\label{tab:tab_rf}
\end{center}
\end{table}

\subsection{XGBoost}

Boosting is a very popular machine-learning technique. It is a type of ensemble learning that uses the output of the previous model as input to the next one. Instead of training models individually, boosting trains models sequentially, with each new model trained to correct the errors of previous ones. At each iteration, correctly predicted results are given less weight, and incorrectly predicted results are given more weight. It then uses the weighted average to get the final result. 

We are also interested in boosting from the point of view that often a certain principle of handling missing values is built into the algorithm. Two popular boosting models are CatBoost \citep{2017arXiv170609516P} and XGBoost \citep{Chen2016}. CatBoost only has built-in filling in the missing values with some specific values, but XGBoost uses the following strategy: each node is assigned a default solution, and this works well in many cases. Therefore, we choose XGBoost for the task and also compare the performance of the model on the missing values filled in by the default method of XGBoost and filling with the Iterative Imputer. 

On a test dataset with the default missing value algorithm, XGBoost gives $\textsc{MCC}=0.96$. When training and testing on data in which we filled the missing values using the Iterative Imputer, the performance reaches $\textsc{MCC}=0.986$. Thus, we conclude that Iterative Imputer in this case is not only a more robust method but also has a better effect on model performance.

The number of estimators is fixed at 500. The following hyperparameters were optimized with optuna: maximum tree depth, learning rate (the step size reduction used during the update to prevent overfitting), subsample ratio of the training instances, reg\_alpha (regularization term) and gamma - the minimum loss reduction required to create a further partition on a leaf node of the tree. Optimized values are presented in Tab.~\ref{tab:tab_xg}.

\begin{table}[ht]
\begin{center}
\caption{XGBoost hyperparameters for three sets of features.}
\scriptsize
\begin{tabular}{lccc}
Hyperparameter     & All features      & No PS magnitudes & Only colours  \\
\hline
\hline
Max depth & 15 & 15 & 10 \\
\noalign{\smallskip}\hline\noalign{\smallskip}
Learning rate & 0.340 & 0.126 & 0.033 \\
\noalign{\smallskip}\hline\noalign{\smallskip}
Subsample & 0.05 & 0.04 & 0.11 \\
\noalign{\smallskip}\hline\noalign{\smallskip}
Gamma & 0.323 & 0.548 & 0.996 \\
\noalign{\smallskip}\hline\noalign{\smallskip}
Reg alpha & 0.82 & 0.48 & 0.03 \\
\hline
\hline
\noalign{\smallskip}\noalign{\smallskip}
MCC test score &  0.978 & 0.972 & 0.969 \\
\noalign{\smallskip}\hline\noalign{\smallskip}
MCC train score & 0.980 & 0.978 & 0.974 \\
\noalign{\smallskip}\hline\noalign{\smallskip}
Precision & 0.992 & 0.989 & 0.985 \\
\noalign{\smallskip}\hline\noalign{\smallskip}
Recall & 0.985 & 0.982 & 0.985 \\
\hline
\hline
\end{tabular}
\label{tab:tab_xg}
\end{center}
\end{table}

The feature importances is presented in Fig.~\ref{fig:importance}. Trained on all features XGBoost relies mainly on the $i$ magnitude of the Pan-STARRS survey, as does RF. If Pan-STARRS magnitudes are excluded from the feature list, $i-z$, $z-J$ and $i-y$ colours became the dominant features, however, the 
$J$ magnitude of the 2MASS survey remains important. 

\subsection{Support Vector Machine}

The support vector machine \citep{cortes1995support} is another widely used and well-developed method. The principle of the support vector machine is to find a line, surface or hypersurface that would separate classes in the feature space. The fitting process maximizes the distance from each point to the decision boundary (reference vector).

\begin{figure}
	\begin{minipage}[ht]{\linewidth}	\center{\includegraphics[width=0.8\textwidth]{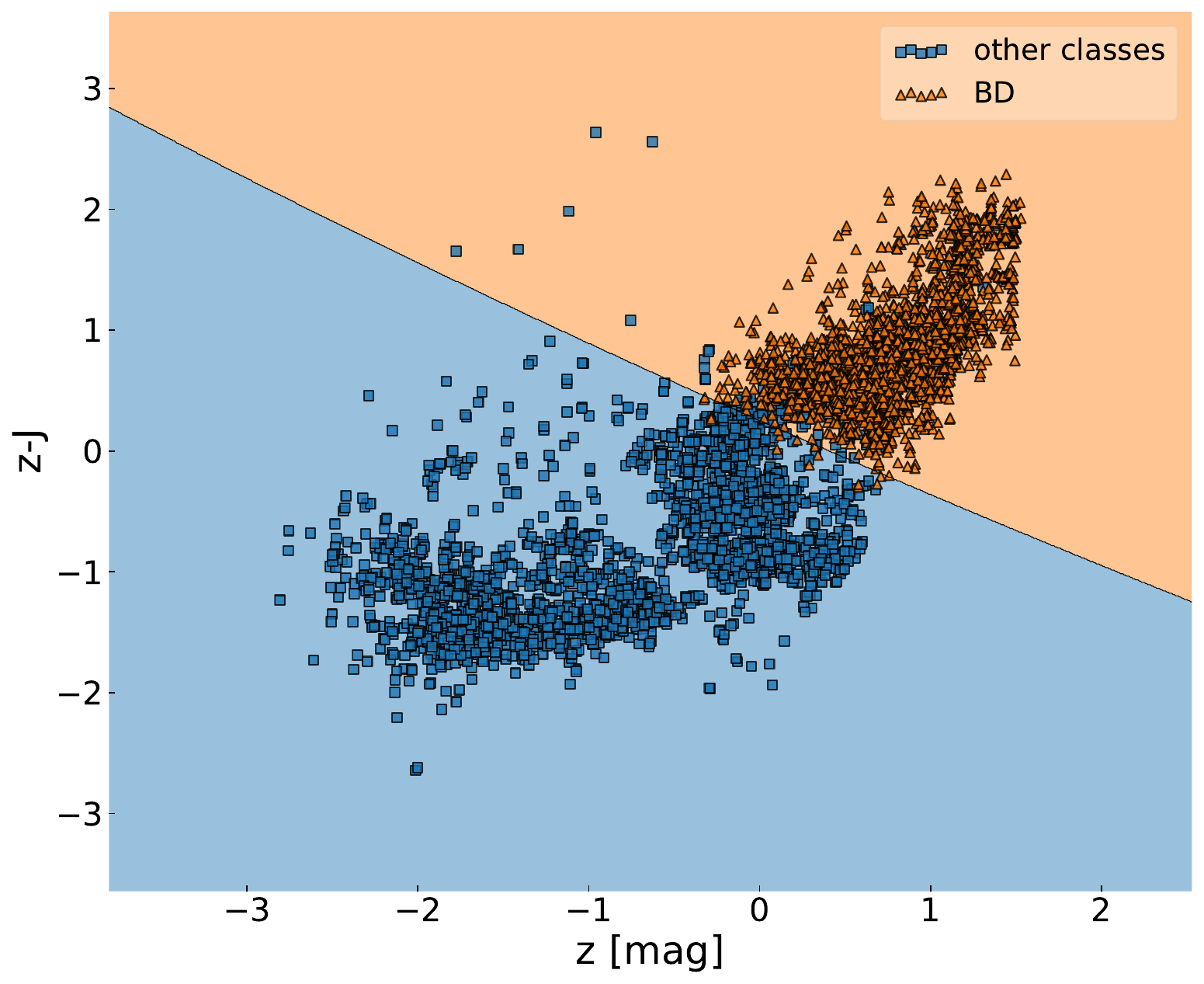}}
	\end{minipage}
    \caption{Slice of the separating boundary in the space of features according to the SVM model.} 
    \label{fig:cut_svm}
\end{figure} 

We tuned the regularization parameter C, the kernel type and the kernel coefficient ('gamma', for 'rbf' kernel). The decision function was set to one-versus-one ('ovo') since it is a binary classification and weights of classes were  automatically adjusted inversely proportional to class frequencies in the input data. For tuned parameters in every case see Tab.~\ref{tab:tab_svc}. It is seen from the table that models are very similar for all of the cases and so are the most important features (see Fig.~\ref{fig:importance}). As SVM mainly rely on colour indices, the importance distributions do not change significantly when some or all of the magnitudes get excluded. 

\begin{table}[ht]
\begin{center}
\caption{SVM classifier hyperparameters for three sets of features.}
\scriptsize
\begin{tabular}{lccc}
Hyperparameter     & All features      & No PS magnitudes & Only colours  \\
\hline
\hline
Kernel & rbf & linear & rbf \\
\noalign{\smallskip}\hline\noalign{\smallskip}
C & 1.150 & 0.729 & 0.792 \\
\noalign{\smallskip}\hline\noalign{\smallskip}
Gamma & 0.298 & 0.066 & 0.453 \\
\hline
\hline
\noalign{\smallskip}\noalign{\smallskip}
MCC test score & 0.981 & 0.984 & 0.958 \\
\noalign{\smallskip}\hline\noalign{\smallskip}
MCC train score & 0.982 & 0.980 & 0.968 \\
\noalign{\smallskip}\hline\noalign{\smallskip}
Precision & 0.989 & 0.986 &  0.972 \\
\noalign{\smallskip}\hline\noalign{\smallskip}
Recall & 0.992 & 0.998 & 0.986 \\
\hline
\hline
\end{tabular}
\label{tab:tab_svc}
\end{center}
\end{table}

Fig.~\ref{fig:cut_svm} shows a separating boundry constructed by the SVM model using a 2D cut as an example. It is worth noting that this is only a slice in which the remaining feature values are taken in some neighbourhood of the mean, so a large number of points that fell into the wrong area does not mean an actual misclassification.

\subsection{TabNet}

TabNet \citep{Arik2019} is a deep learning neural network that utilises attention to select important features at each step of the decision-making process so that only the most important features are used. In this case, the choice of features depends on the object, and, for example, it can be different for each row of the training data set. In the end, you can see which features the model has focused on the most.

TabNet consists of several steps, each step is a block of components, with the number of steps being a hyperparameter. Each step gets its vote in the final classification, which mimics the ensemble classification. Other hyperparameters are the width of the decision prediction layer (N\_d), the width of the attention embedding for each mask (N\_a), number of shared Gated Linear Units at each step (N\_shared) and the coefficient for feature reusage in the masks (Gamma). We fitted the hyperparameters of TabNet on optuna using MCC score as the metric. The fitted parameters are presented in Tab.~\ref{tab:tab_tbn}.

\begin{table}[ht]
\begin{center}
\caption{TabNet hyperparameters for the different sets of features.}
\scriptsize
\begin{tabular}{lccc}
Hyperparameter     & All features      & No PS magnitudes & Only colours  \\
\hline
\hline

N\_d & 16 & 12 & 12  \\
\noalign{\smallskip}\hline\noalign{\smallskip}
N\_a & 16 & 28 & 12 \\
\noalign{\smallskip}\hline\noalign{\smallskip}
Number of steps & 3 & 3 & 2 \\
\noalign{\smallskip}\hline\noalign{\smallskip}
Gamma & 1.2 & 1.6 & 1.6 \\
\noalign{\smallskip}\hline\noalign{\smallskip}
N\_shared & 2 & 2 & 1 \\
\hline
\hline
\noalign{\smallskip}\noalign{\smallskip}
MCC test score & 0.983 & 0.986 & 0.975 \\
\noalign{\smallskip}\hline\noalign{\smallskip}
MCC train score & 0.987 & 0.990 & 0.990 \\
\noalign{\smallskip}\hline\noalign{\smallskip}
Precision & 0.992 & 0.992 & 0.988 \\
\noalign{\smallskip}\hline\noalign{\smallskip}
Recall & 0.992 & 0.994 & 0.987 \\
\hline
\hline
\end{tabular}
\label{tab:tab_tbn}
\end{center}
\end{table}

\begin{figure}
	\begin{minipage}[ht]{\linewidth}
		\center{\includegraphics[width=0.8\textwidth]{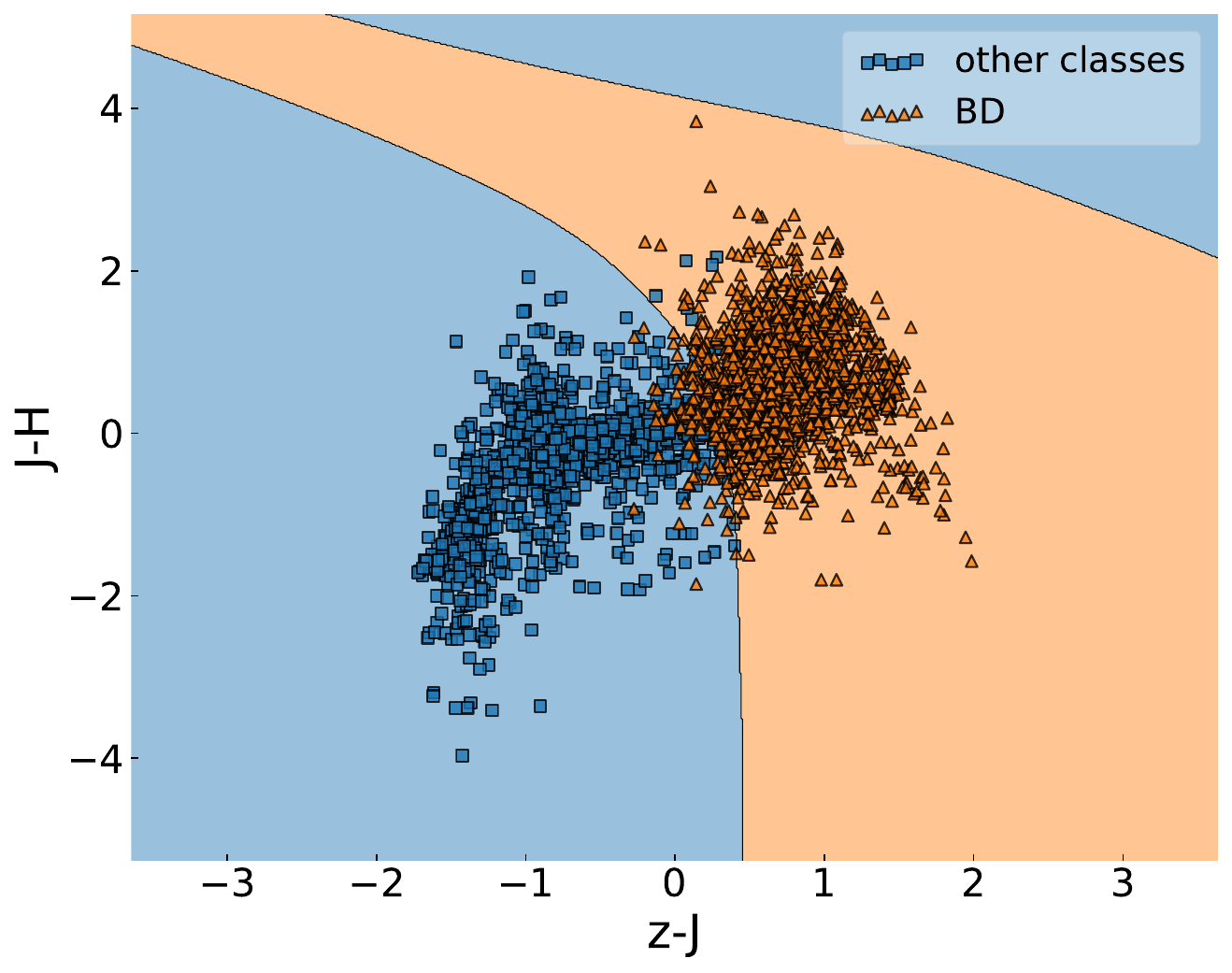}}
	\end{minipage}

    \caption{Slice of the separating boundary in the space of features according to the TabNet model.} 
    \label{fig:cm_tbn}
\end{figure}

The model is trained using gradient descent optimization with Adam as the optimizer. The validation part of a dataset is used in training 
to prevent overfitting, so the result of the model training is the configuration, that provides the best scores, both on training and validation data. A separating boundary is shown in Fig.~\ref{fig:cm_tbn}, as it can be seen it is more complex then the boundaries of other models.     

\section{Results and discussion}
\label{sec:res}

In this section, we summarise the results of the application of machine learning methods to the brown dwarf search problem. 

We trained four models on the data: Random Forest Classifier, SVM Classifier, XGBoost Classifier and TabNet Classifier. 

\begin{figure*}[ht]
    \centering
    \includegraphics[width=0.95\linewidth]{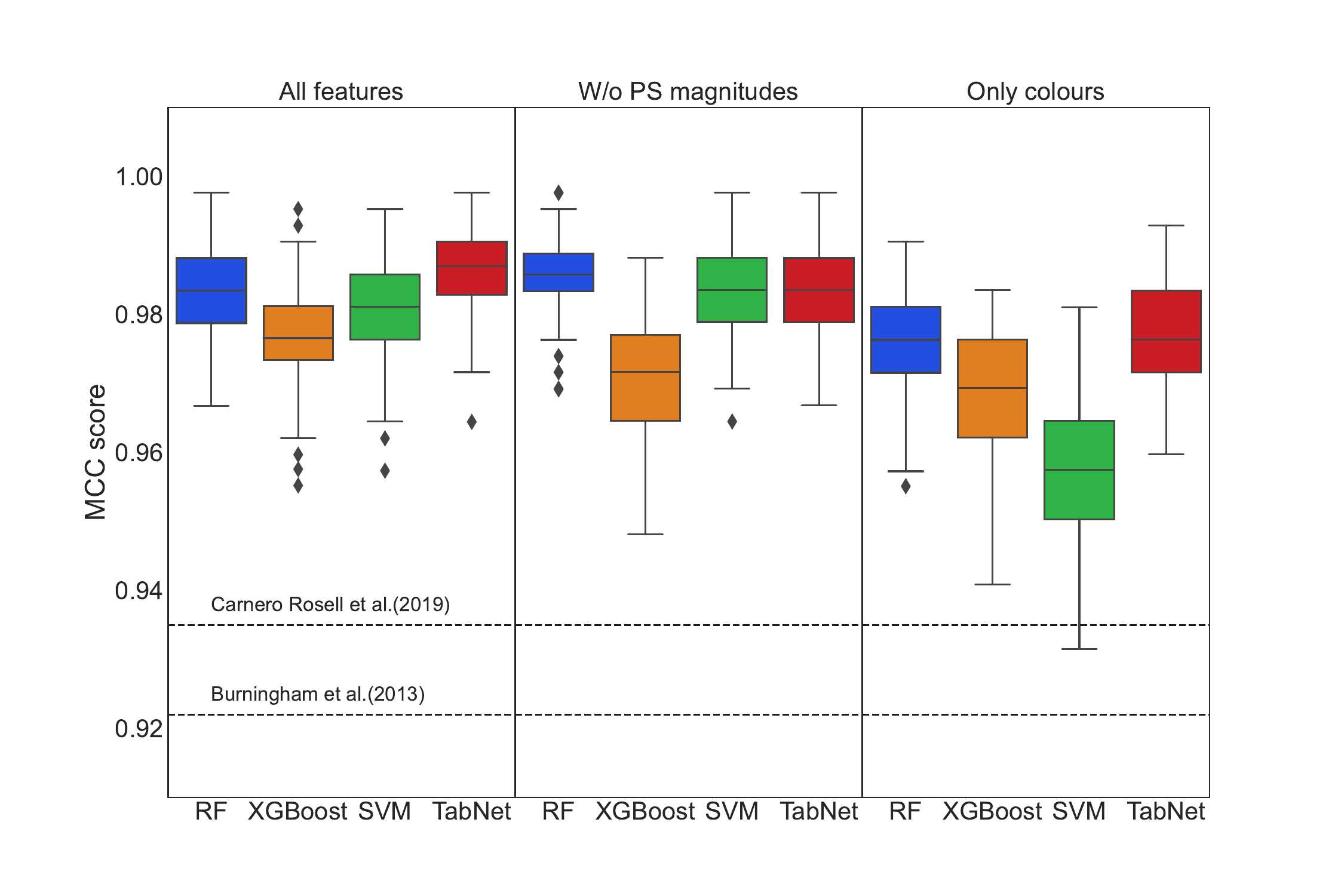} 
    \caption{Confidence intervals for the scores of models}
    \label{fig:trust}
\end{figure*}

Confidence intervals for the MCC scores of the obtained models are presented in  Fig.~\ref{fig:trust}. Confidence intervals are calculated via the bootstrap method with 100 samples half the length of the test data set. The coloured box is the interval from the 25th percentile to the 75th percentile and the median value is represented by a black line. Error bars show the minimum and the maximum value with the outliers marked as diamonds. The baseline values obtained using decision rules from the literature are represented in Fig.~\ref{fig:trust} as dashed lines.

All models on the full set of features and the features without PS1 magnitudes provide almost the same results, but some performed slightly better. If we use only magnitude features, the performance decreases. However, such models still outperform the baselines. 

\begin{figure*}[ht]
    \centering
    \includegraphics[width=\linewidth]{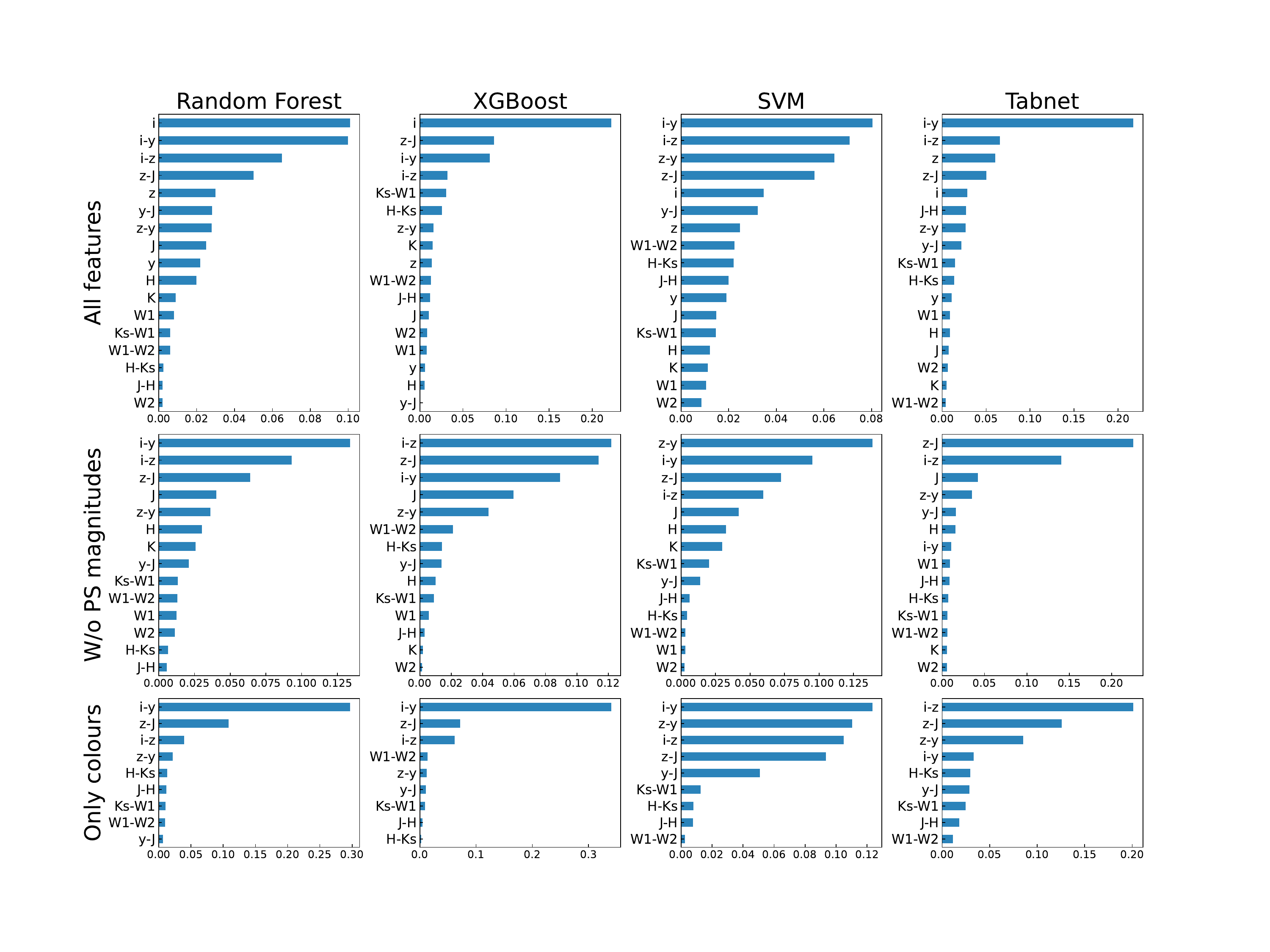} 
    \caption{Importance of features for all models. For RF, XGBoost and SVM models we calculate the importance of each feature using \textit{SHAP} and for TabNet we use  built-in capabilities of the model.}
        \label{fig:importance}
\end{figure*}

Although the performance of the models is nearly the same, they differ in terms of robustness. The importance of features of the models is presented in Fig.~\ref{fig:importance}. Random Forest and XGBoost rely primarily on the $i$ magnitude in their decisions, if any. At the same time, SVM and Tabnet on the full set of features seem more robust, as they rely mainly on colour indices. Tabnet also tends to pay a lot of attention to the magnitudes, although it performs the best in the case of using only colours. Table~\ref{tab:confusionallcases} shows true positive, true negative, false positive and false negative values for all of the models obtained on the test part of the dataset. 

According to \cite{CarneroRosell2019} and \cite{Skrzypek2015}, $(i-z)_{PS1}$ and $y_{PS1}-J$ colour indices are expected to be the most important feature since they have the largest variation across the M/L transition. While the $(i-z)_{PS1}$ colour index is important to SVM classifiers and in some cases for RF and TabNet, most of the models do not consider them essential. 

It is expected that $z_{PS1}-J$ is the most important feature as \cite{Burningham2013}, which we used as a baseline, is nearly entirely based on the above colour. However, it only plays a secondary role in most of the models. Unlike the previous works, we revealed the importance of the $(i-y)_{PS1}$ colour index. It is the most important feature in most cases and the colour selection $(i-y)_{PS1}>1.88$ alone gives an MCC score of $0.968$ on the testing data. Other colour indices could be important in the case of multi-class classification, for example, L and T-type dwarfs differ drastically in $W1-W2$, although it has almost no relevance for the problem investigated in the present article.

\begin{table*}[ht]
\begin{center}
\caption{True positive (TP), true negative (TN), false positive (FP) and false negative (FN) values as well as Precison and Recall scores for four models: Random Forest (RF), XGBoost, Support Vector Machine (SVM), and TabNet. Each model was trained on three sets of features, labeled as "All features", "w/o PS magnitudes", and "only colours".}
\begin{tabular}{c|cccccc}
Model     & TP      &  TN & FP & FN & Precision & Recall      \\
\hline
\hline

 \multicolumn{7}{c}{Random Forest}  \\
\hline
All features & 846 & 7 & 846 & 7 & 0.992 & 0.992 \\
W/o PS magnitudes & 848 & 7 & 833 & 5 & 0.992 & 0.994 \\
Only colours & 842 & 10 & 830 & 11 & 0.988 & 0.987 \\

\hline
\hline

 \multicolumn{7}{c}{XGBoost}  \\
\hline
All features & 840 & 6 & 834 & 13 & 0.992 & 0.985 \\
W/o PS magnitudes & 838 & 9 & 831 & 15 & 0.989 & 0.982 \\
Only colours & 840 & 13 & 827 & 13 & 0.985 & 0.985\\

\hline
\hline

 \multicolumn{7}{c}{SVM}  \\
\hline
All features & 846 & 9 & 831 & 7 & 0.989 & 0.992 \\
W/o PS magnitudes & 851 & 12 & 828 & 2 & 0.986 & 0.998 \\
Only colours & 841 & 24 & 816 & 12 & 0.972 & 0.986 \\

\hline
\hline

 \multicolumn{7}{c}{TabNet}  \\
\hline
All features & 850 & 9 & 831 & 3 & 0.992 & 0.992 \\
W/o PS magnitudes & 851 & 12 & 828 & 2 & 0.992 & 0.994 \\
Only colours & 846 & 13 & 827 & 7 & 0.988 & 0.987 \\
\hline

\end{tabular}
\label{tab:confusionallcases}
\end{center}
\end{table*}

\section{Conclusion}

In this paper, we compiled a dataset of L and T type brown dwarfs (marked as positive class) and objects of other spectral types (marked as negative class) from literature. It is important to acknowledge potential limitations and biases that could impact the dataset's representativeness and generalizability. While we have compiled the data set in such a way as to reproduce the observed distribution in absolute magnitudes, the distribution in apparent magnitudes may be not representative, shifted towards brighter objects. This most likely does not affect the performance of the model in the case of "only colours" mode, but may have an impact on models trained on magnitudes too. The incompleteness in M type dwarfs that stems from limitations in the \cite{Best2018} catalogue introduces a bias in the dataset towards the included objects, potentially impacting the generalizability of the results.

Brown dwarfs are faint astronomical sources with peak intensities that fall into the infra-red part of the spectrum, so it is harder to carry out the measurements in optical and far-red filters, such as $i$ and $z$. We imputed missing values with Iterative Imputer and explored the result. We also imputed the colour indices regardless of corresponding magnitudes in order to reduce the error of imputing. For most of the magnitude features, the imputing error is compatible with the measurement error. For colour index features, imputing error is usually much lower than the corresponding measurement error. Thus, the imputing part of preprocessing is considered to be successful. 

Four models, namely, Random Forest Classifier, SVM Classifier, XGBoost Classifier and TabNet Classifier were trained to distinguish brown dwarfs among all objects according to their photometric data. The classification results for all models are consistently high, all models outperform the baselines. However, tree-like models (RF and XGBoost) tend to exploit the faintness of the objects, which is less preferable in terms of robustness of the model. On the contrary, SVM primarily relies on colour indices, which reduce the possibility of misclassification of other types of faint sources. 

Examining the features of the models, we have found that the $(i-y)_{PS1}$ colour index is the most important feature in most cases. This colour index can be used subsequently and independently in brown dwarf classification problems. We also confirm that the $z_{PS1}-J$, $(i-z)_{PS1}$ and $(z-y)_{PS1}$ colour indices are important and powerful features in brown dwarf classification. 

Along with the overall success of the work, there are several limitations to the applicability of the results, that are yet to be solved. First, potential biases of the dataset were emphasised, including a bias towards brighter objects. These biases can affect the generalizability and reliability of the findings. Second, it is known, that highly red-shifted quasars might also be a source of contamination when distinguishing brown dwarfs from other types of objects \citep{CarneroRosell2019}. This issue can be resolved using colour cuts in most cases, but can still remain for some cases \citep{Reed2017}. At this point, we haven't considered quasars. Third, while the imputing magnitudes and colour indices independently allow us to predict colour indices much more accurately, this leads to the colour index not being always a linear combination of magnitudes involved. Multi-class classification will also be in scope of future work.  

\section*{Acknowledgements}

This work was supported by by Non-commercial Foundation for the Advancement of Science and Education INTELLECT.

Author is grateful to Ilya Dyugay, Konstantin Malanchev, Dana Kovaleva and Oleg Malkov for the fruitful discussion and valuable advice. Author thanks the anonymous referee for careful reading and suggestions, which greatly helped to improve the article. 

This research has made use of NASA’s
Astrophysics Data System Bibliographic Services.

\bibliographystyle{model1-num-names}

\bibliography{nn}

\end{document}